\documentclass[graphicx,11pt]{aastex}

\lefthead{}
\righthead{}

\begin{document}

\title{Origin of the peculiar eccentricity distribution of the inner cold Kuiper belt} 

\author{\textbf{A. Morbidelli$^{(1)}$,H. S. Gaspar $^{(2)}$,
    D. Nesvorny$^{(3)}$}\\  
(1) Laboratoire Lagrange, UMR7293, Universit\'e de Nice Sophia-Antipolis,
  CNRS, Observatoire de la C\^ote d'Azur. Boulevard de l'Observatoire,
  06304 Nice Cedex 4, France. (Email: morby@oca.eu / Fax:
  +33-4-92003118) \\
(2) UNESP – Univ. Estadual Paulista, FEG. GDOP, Av. Dr. Ariberto Pereira da Cunha, 333, CEP 12\,516-410, Guaratinguet\'a, SP, Brazil (Email: helton.unesp@gmail.com); Capes Foundation, Ministry of Education of Brazil, Caixa Postal 250, Bras{\'\i}lia – DF 70040-020, Brazil \\ 
(3) Southwest Research Institute, Boulder, Co. \\ 
} 

\begin{abstract}

Dawson and Murray-Clay (2012) pointed out that the inner part of the
cold population in the Kuiper belt (that with semi major axis
$a<43.5$~AU) has orbital eccentricities significantly smaller than the
limit imposed by stability constraints. Here, we confirm their result
by looking at the orbital distribution and stability properties in
proper element space. We show that the observed distribution could
have been produced by the slow sweeping of the 4/7 mean motion
resonance with Neptune that accompanied the end of Neptune's migration
process. {The orbital distribution of the hot Kuiper belt
  is not significantly affected in this process, for the reasons discussed in the main text.} Therefore, the
peculiar eccentricity distribution of the inner cold population can
not be unequivocally interpreted as evidence that the cold population
formed in-situ and was only moderately excited in
eccentricity; it can simply be the signature of Neptune's
radial motion, starting from a moderately eccentric orbit. We discuss
how this agrees with a scenario of giant planet evolution following a
dynamical instability and, possibly, with the radial transport of the
cold population.

\end{abstract}

\section{Introduction}

The Kuiper belt has a complex orbital structure and can be divided in
multiple sub-populations (see Gladman et al., 2008 for a
review). Among them are the cold and the hot populations, which are
defined as the collections of objects inwards of the 1/2 resonance
with Neptune ($\sim 48$~AU) with, respectively, inclinations smaller or 
larger than 4 degrees. The
cold and hot populations have also distinct physical properties
(see Morbidelli and Brown, 2004 for a review).

There is a quite general consensus that the hot population formed
originally closer to the Sun, was dynamically excited by the
perturbations from the giant planets and finally was transported into
the Kuiper belt (Gomes, 2003; Levison et al., 2008). However, there is
no consensus on the origin of the cold population. Some models argue
that it also was transported into the Kuiper belt from a region closer
to the Sun (Levison and Morbidelli, 2003; Levison et al., 2008), while
others argue that the cold population formed locally (e.g. Parker et
al., 2011; Batygin et al., 2011).

An important point in this debate was made by Dawson and Murray-Clay
(2012).  First they observed that the usual partition of the cold and
hot populations according to the 4-degree inclination boundary is
simplistic; in reality these populations have distinct, but partially
overlapping inclination distributions (see Brown, 2001). Thus, to
limit the contamination of the cold population by the hot population,
they restricted their analysis to objects with inclination
$i<2^\circ$, where the relative fraction of low-inclination ``hot''
objects is expected to be negligible. Then they showed that, inside of
43.5~AU, this low-inclination population has also small eccentricities
($e\lesssim 0.05$), even though orbits would be stable up to $e\sim
0.1$ (Lykawka and Mukai, 2005a).  The hot population, in fact, has
eccentricities up to this limit.  The lack of moderate eccentricity
orbits in the cold population obviously cannot be explained by
observational biases. Dawson and Murray-Clay therefore interpreted this result
as evidence that the cold Kuiper belt was only very moderately excited
relative to its original quasi-circular and coplanar orbits. 
This argues against models in which the cold population originates closer to the sun and is implanted into the Kuiper belt, because such models predict a cold population with an eccentricity distributions covering the whole stability range.

Given the importance of this argument, we have decided to revisit the
problem of the eccentricity distribution of the cold Kuiper belt. In
Section 2 we redo the same analysis as Dawson and Murray-Clay, but
using proper elements instead of osculating elements. Our results
confirm theirs, but {we notice that} the transition between the
inner part of the cold population, where eccentricities are all small,
to the outer part, where the eccentricities cover a wider range,
happens exactly at the 4/7 mean motion resonance with Neptune. This
suggests that this resonance might have played a role in sculpting the
inner cold belt during a phase of outward migration. Then, in Section
3 we conduct migration experiments, testing different migration
timescales and eccentricities of Neptune. {Section 4 analyzes more
  in details how moderate-eccentricity cold Kuiper belt objects are
  removed by resonance sweeping and compares their evolution with that of high inclination bodies.} Our conclusions are
  discussed in Section 5.

\section{Distribution of proper elements and stability map for the cold Kuiper belt}

We have selected all TNOs {from the MPC catalog} with semi major axis larger than 25 AU and
orbits determined from observations covering at least three
oppositions. Such procedure selected a set of 811 TNOs.  For each of
these objects we computed numerically the orbital proper elements
using integrations covering 132~My.  The proper semi major axis was
computed by numerical average of the values recorded during the
simulation, with an output time-step of 1,000y. For the proper
eccentricities and inclinations, the computational procedure was more
elaborated, although standard (Knezevic and Milani, 2000).  We first
computed the Fourier series of $(h(t), k(t))$ and $(p(t), q(t))$,
where:
\begin{eqnarray}
h(t)&=&e(t)\cos[\varpi(t)]\cr
k(t))&=&e(t)\sin[\varpi(t)]\cr
p(t)&=&i(t)\cos[\Omega(t)]\cr
q(t))&=&i(t)\sin[\Omega(t)]
\end{eqnarray}
and $e(t), i(t), \varpi(t)$ and $\Omega(t)$ are the values of
eccentricity, inclination, longitude of perihelion and longitude of
node recorded over time $t$. We then removed from the series
expansions the terms with frequencies close (i.e. within one arcsec/y)
to the proper frequencies of the planets. Finally, we selected as
proper eccentricity and inclination the coefficients of the largest
remaining term in the each of the two Fourier series.

In order to evaluate the accuracy of proper eccentricity and
inclination, we performed the procedure described above for the
first and the second halves of the whole integration, i.e.,
65.536\,My. Then, we adopted as an estimate of the error, the largest
difference among the proper elements calculated for the whole
integration time-span and those computed in each of the two
half time-spans.  The relative accuracy in proper semi major axis was
always better than $3\times 10^{-4}$.  The absolute accuracies in
proper eccentricity and inclinations were better than 0.01 and 0.1
degrees throughout the region of interest (i.e. inside of 44 AU and
not in the 4/7 mean motion resonance with Neptune).

\begin{figure}[t!]
\centerline{\includegraphics[height=8.cm]{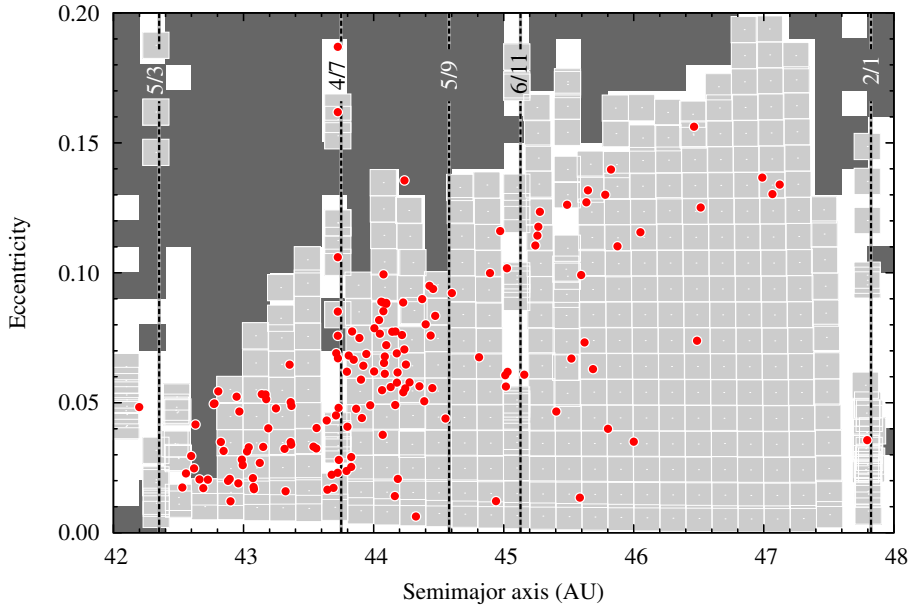}}
\caption{\small The dots show the distribution of proper semi major
  axis and proper eccentricities of the Kuiper belt objects with
  well-defined orbits and proper inclination smaller than 2
  degrees. The light-gray squares denote the regions of proper element
  space that are stable in 4~Gy simulations and the dark-gray squares
  denote the initial orbital elements of unstable particles. White
  color is the background.  The light-gray squares are not regularly
  spaced because the application mapping the
  initial conditions (regularly spaced) to proper elements is not
  linear. The vertical dashed lines depict the main mean motion
  resonances as labeled.  }
\label{Helton}
\end{figure}

Once in possession of this proper element catalog, following Dawson and
Murray-Clay we retained as members of an ``uncontaminated cold population'' the
objects with proper inclination smaller than 2
degrees. Fig.~\ref{Helton} shows the distribution of the selected
objects (dots) in proper semi major axis vs. eccentricity plane.

We also computed a new stability map. In fact, the map used by Dawson
and Murray-Clay, from Lykawka and Mukai (2005a), was computed for a wide
range of inclinations, whereas here we are interested to very low
inclinations only. We could have used the stability map in Duncan et
al. (1995), which was computed for $i=1^\circ$, but the latter was
quite sparse, due to the computing limitations of the time. Moreover,
both Lykawka and Mukai and Duncan et al. reported their maps relative
to the initial osculating elements. Here, for a consistent comparison
with the proper elements of the real objects, we needed a map computed
in proper elements space.

To compute the stability map, we proceeded as follows. We adopted a
grid of particles' initial conditions, with {osculating elements in the following ranges:} 42~AU~$< a <$~48~AU, $0 <
e < 0.2$ and $0 < i < 2$ deg, with resolutions of 0.2~AU in $a$, 0.01
in $e$ and 0.5 degrees in $i$. The secular angles $\Omega$ and
$\varpi$ were set equal to 0 degrees.  Each particle was
integrated for 132~My. We then computed their proper elements following
the same procedure described above. Finally, we continued the
simulations for 4~Gy in order to asses the long-term survival
of the test particles.

To construct Fig.~\ref{Helton}, for each given pair of initial $a$ and
$e$, we selected the particle with the smallest value of
proper inclination. Then, for the particles that survived in the 4~Gy
integration (i.e. they did not encounter Neptune within a Hill radius
within this time), 
we plotted on a white background a light-gray square of
size 0.2~AU$\times 0.01$ centered on their values of proper semi major
axis and eccentricity measured on the first 132~My. 
Moreover, particles 
that did not survive 
were denoted by dark-gray squares centered 
on the initial pair of osculating semi major axis and eccentricity.

The stability map of Fig.~\ref{Helton} shows few surprises. In
general, particles are unstable at large eccentricity and stable at
low eccentricity, where the perihelion distance is larger than $\sim
38$--40~AU. Mean motion resonances represent the exception to this
general rule.  The light-gray squares at large eccentricity are all
associated to mean motion resonances, as well as the vertical white
columns at low or moderate eccentricities. Some mean motion
resonances, therefore, clearly stand out from the stability map, and
they are labelled on the figure.

In general, as expected, the dots fall on light-gray squares. Those that
don't, are associated to mean motion resonances. In fact, in a mean
motion resonance there is a third dimension characterizing the orbit:
the resonant amplitude. It is well possible that none of the test
particles that we used for the stability map sampled the orbit of a
real particle because their libration amplitude is different.  In this
case, a dot is plotted over the white background.

Overall, Fig.~\ref{Helton} confirms the results of Dawson and
Murray-Clay.  Inside of 43.5 AU, all real objects have small
eccentricities, barely exceeding 0.05. The stability map, however,
ranges up to 0.1 in the 42.2-42.6~AU region. Thus, there is clearly a
stable region (approximately in the range $0.05<e<0.1$) in the inner
belt that is not inhabited by the cold population. At a
closer inspection, one sees that the transition in eccentricity
distribution of the cold belt is sharply at the location of the 4/7
resonance (see Lykawka and Mukai, 2005b for a description of this
resonance). 

This result suggests that the 4/7 resonance might have played a role
in depleting the moderate-eccentricity cold objects inside of its
current location, as it swept through the 42.5--43.5~AU region during
the putative radial migration of Neptune. Thus, in the following
section, we report on numerical experiments of radial migration that address
whether this is indeed possible and at which conditions.

\section{Migration numerical experiments: the role of the 4/7 resonance sweeping}

We set up simple numerical experiments, where Jupiter, Saturn and
Uranus were assumed to be on their current orbits, while Neptune's mean
semi major axis was forced to change from $a_i=28.5$~AU to its current
average value $a_c=30.1$AU as
$$a(t)=a_c+(a_i-a_c)\times\exp(-t/\tau)$$. 
The eccentricity of Neptune was also damped from an initial value $e_0$ as 
$$\frac{1}{e} \frac{{\rm d}e}{{\rm d}t}= \alpha\exp(-t/\tau)$$ with an
appropriate coefficient $\alpha$ that allowed us to match the current
eccentricity of Neptune at the end of the simulation. Each simulation
was run for a time-span of $6\times\tau$. The semi major axis drift and
the eccentricity damping were implemented by applying synthetic forces
to Neptune's equations of motion. In particular, we employed the forces
described in Malhotra (1995) for evolving the semi-major axes and
those in Kominami et al. (2005) for controlling the eccentricity.

\begin{figure}[t!]
\centerline{\includegraphics[height=12.cm]{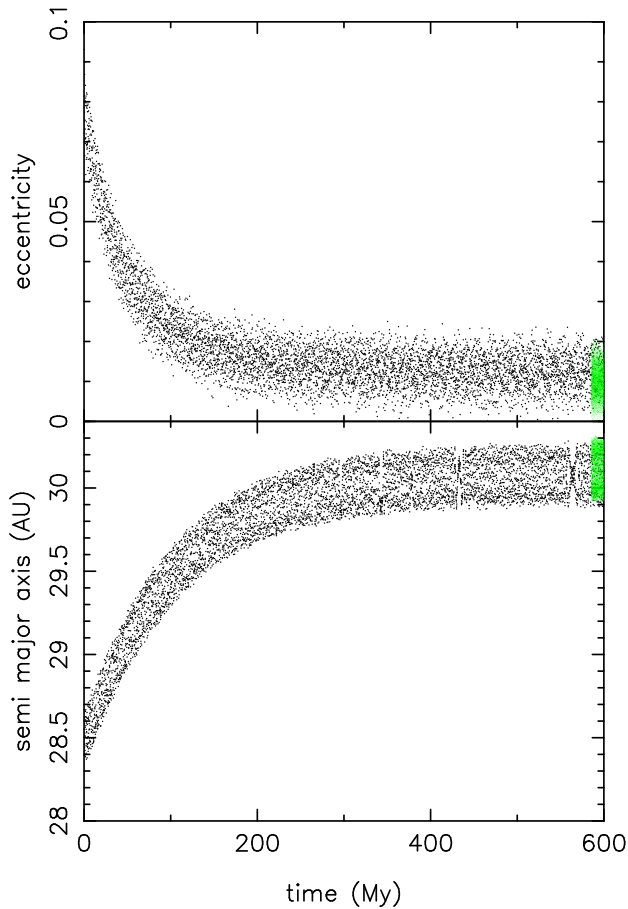}}
\caption{\small The evolution of Neptune during the migration
  experiment with $\tau=100$~my and $e_0=0.1$. The top panel depicts
  the time evolution of the eccentricity and the bottom panel that of
  the semi major axis. The green area close to the $t=600My$ axis is
  obtained by plotting the evolution of Neptune in the current solar
  system, and represents the target $a$ and $e$ 
  that a good migration simulation needs to hit.}
\label{Neptune}
\end{figure}

Fig.~\ref{Neptune} shows the evolution of Neptune in a simulation with
$\tau=100$~My and $e_0=0.1$. Notice that, although the {\it initial}
eccentricity is 0.1, the {\it mean} eccentricity at the beginning of
the simulation is only 0.075. The green area at the right-hand-side of
each panel shows the current range of oscillations of the semi major
axis and eccentricity of Neptune in the real Solar System. This shows
that our migration evolution reproduces the actual orbit of Neptune
with a good accuracy.
  
Overall we did 9 simulations, corresponding to the values of $\tau$
and $e_0$ reported in Table~\ref{table}. {These values cover a range of
proposed migration models, with and without a putative initial eccentricity excitation of Neptune's orbit. We will discuss in Sect.~\ref{conclusions} what the successful parameters imply for the history of the Solar System.} 
In each simulation we
considered a population of 1,000 test particles, initially spread
{in osculating elements} between 42.3 and 44.2~AU (in some cases only up to 43.6 AU) in semi major axis, from 0 to 0.15 in
eccentricity and up to 2 degrees in inclination (with a uniform
distribution in $\sin(i)$). Particles got discarded when they had
encounters with Neptune. For the surviving particles we computed the
proper elements over the final part of the simulation. Only particles
with final proper inclination smaller than 2 degrees were considered.

\begin{figure}[t!]
\centerline{\includegraphics[height=9.cm]{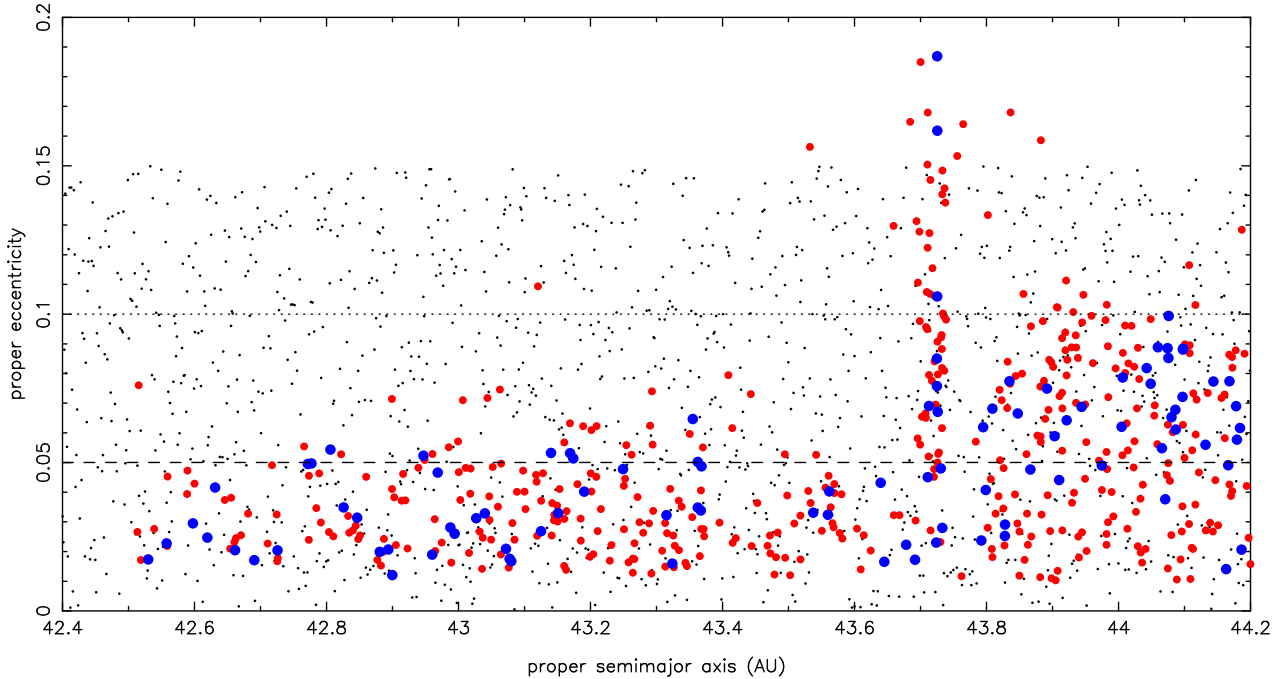}}
\caption{\small Comparison between the real distribution and the
  simulated distribution at the end of the simulation with
  $\tau=100$~My and $e_0=0.1$. The red dots depict the simulated
  particles and the blue dots the real cold belt objects (the same as
  shown in Fig.~\ref{Helton}). The column of objects at $a\sim
  43.72$~AU marks the location of the 4/7 resonance with Neptune. The
  small black dots depict the initial conditions of the simulated
  particles.  Horizontal lines at $e=0.1$ and $e=0.05$ are plotted for
  reference.  }
\label{results}
\end{figure}

Fig.~\ref{results} shows the final particle distribution (red dots)
obtained in the simulation with $\tau=100$~My and $e_0=0.1$. Their
final proper semi major axes have been multiplied by a factor 1.00092,
to compensate for the slight offset of the final semi major axis of
Neptune relative to the real one, visible in Fig.~\ref{Neptune}. For
comparison, the blue dots represent the real low-inclination cold
population (the same as in Fig.~\ref{Helton}) and the small black dots
show the initial conditions.

First, notice the cluster of simulated particles in the 4/7 resonance
at $\sim 43.72 AU$. This cluster is visible in the observations as
well, in the right proportion. In fact, the ratio between the number
of resonant particles and that of particles with $a<43.6$~AU is 26.3\%
at the end of the simulation; among observed objects, this ratio is
25.6\%. {Obviously, one has to take into account that observational biases may act differently on resonant and non-resonant objects, so that this agreement may just be accidental. However, we argue that in this case the differential bias effect is probably not a big issue. In fact, both resonant and non-resonant objects considered here have small inclinations and eccentricities; moreover objects in the 4/7 resonance can reach perihelion at 4 different position in the sky relative to Neptune, so that it is unlikely that all these sweet-spots have been missed by the surveys.}

Second, notice that inside of the final location of the 4/7 resonance,
most of the particles with moderate eccentricities have been removed.
Thus, the final distribution is strongly skewed towards small
eccentricities, more or less similar to the observed distribution.

\begin{table}[!t]
 \caption{A summary of the results of the statistical tests for 9
   simulations.  The first line reports the value of $\tau$ and the
   first column that of $e_0$ (in parenthesis the initial mean value
   for $e$). Each case of the matrix reports $P_1 / P_2 / P_{1+2}$,
   where $P_1$ is the probability that criterion 1 is fulfilled, $P_2$
   is the probability for criterion 2 and $P_{1+2}$ is the probability
   that both criteria are fulfilled simultaneously. See text for
   definition of criteria.}\vskip 10pt \centering
\begin{tabular}[center]{|c|c|c|c|}
  \hline
  ${}_{e_0}\setminus{}^{\tau}$ & 100 My & 30 My & 10 My \\
  \hline
  0.05 (0.036) & 0.035 / 0.18 / 0.015  & 0.0004 / 0.0099 / 0.0001   &  0.0008 / 0.018 / 0.0005    \\
  \hline
  0.1  (0.075)&  0.18 / 0.80 / 0.17   & 0.0019 / 0.0062 / 0.0003  & 0.0 / 0.0008 / 0.0      \\
  \hline
  0.15  (0.14) & unstable   &  unstable  &    unstable   \\
  \hline
\end{tabular}
\vskip 20pt
\label{table}
\end{table}

{Notice also that, beyond 43.8 AU, a similar truncation in the particle distribution occurs, but at proper $e=0.1$. Remember that Fig.~\ref{Helton} shows that, in the current Solar System, orbits in this region would be stable up to $e\sim 0.14$. The truncation at $e=0.1$ is also visible in the observed distribution and in our simulation is operated by the 5/9 resonance sweeping. Resonant sweeping, however, can not explain the ``wedge'', namely the paucity of low eccentricity objects ($e<0.05$) beyond 43.8~AU; therefore, some other explanation is needed (e.g. Batygin et al., 2011) for this structure.}

In order to quantify how well the simulated distribution reproduces
the observed distribution we proceeded as follows. We considered objects
and test particles in the 42.4--43.6 range only. The observed
distribution in this range is made of 43 objects.  The object with the
largest proper eccentricity has $e=0.0653$\footnote{This object is
  1999DA. It was not included in the original analysis of Dawson and
  Murray-Clay, because its current inclination is slightly larger than
  2 degrees. However, its proper inclination is $1.16^\circ$ so it is
  included here}. In total, there are only 7 objects with proper
$e>0.05$.  The simulated distribution has 194 {test particles}. In
this count, we discarded the two particles with proper eccentricity
larger than 0.1 because, from the stability map in Fig.~\ref{Helton},
we know that these particles are unstable on the long term. We then did
a Monte Carlo simulation. From the simulated distribution, we generated
10,000 synthetic populations, each of which contained 43 particles (the
same number as that of the objects in the observed population). We
considered two criteria of ``success''. The first was that a synthetic
population contained no particles with proper eccentricity larger than
0.0653 (criterion 1); the second was that the population contained no
more than 7 particles with proper $e>0.05$ (criterion 2). We found that
18\% of the synthetic populations fulfilled criterion 1; 80\% of them
fulfilled criterion 2 and 17\% fulfilled simultaneously criterion 1 and
2. From this test, therefore, we conclude that the simulated
population is statistically equivalent to the observed population (in
the sense that it cannot be rejected even at a $1-\sigma$
level). Thus, we conclude that the slow migration of the 4/7 resonance
with an initial moderate eccentricity of Neptune can explain the
properties of the inner part of the cold population of the Kuiper
Belt.

The results of the same statistical tests for the other simulations
are reported in Table~\ref{table}. As one can appreciate, fast
migrations with $\tau=10$ or 30~My do a poor job in reproducing the
observations. The simulation with $\tau=100My$ and $e_0=0.05$ does a
decent job if one considers criterion 2, but it is consistent with the
observed population only at the $2-\sigma$ level if one considers
criterion 1. If one considers both criteria simultaneously, the
simulated population has only a 1\% chance to match the observed
population. So, both the migration speed and the initial eccentricity
of Neptune are important (see Sect.~\ref{comparison} for an
explanation of this result).  However, if Neptune is too eccentric
(e.g. in the simulations with $e_0=0.15$) the entire inner Kuiper belt
is destabilized, as described in Levison et al. (2008) and almost no
particles survive with proper $i<2^\circ$ in the considered semi major
axis range.

All these results do not depend strictly on the adopted $i<2^\circ$
limit; they would basically be the same for any reasonable boundary in
proper inclination.

\section{Comparative evolutions of the Hot and Cold populations}
\label{comparison}

{Our proposed explanation for the peculiar orbital structure of the 
cold population would not be acceptable without showing that the hot population avoids being sculpted in a similar fashion by the same process. In fact, as shown by Dawson and Murray-Clay (2012), the hot population does not exhibit any apparent deficit of objects with $e>0.05$.

For this purpose, we run again our best-case simulation, that with  $\tau=100$~My and $e_0$=0.1, with a population of test particles having the following initial osculating orbital elements: $2^\circ < i < 30^\circ$, 42.3 AU $< a <$ 43.6 AU, and $e<0.15$. The initial conditions and the proper elements of the surviving population are shown in Fig.~\ref{Hot}. 
 
\begin{figure}[t!]
\centerline{\includegraphics[height=9.cm]{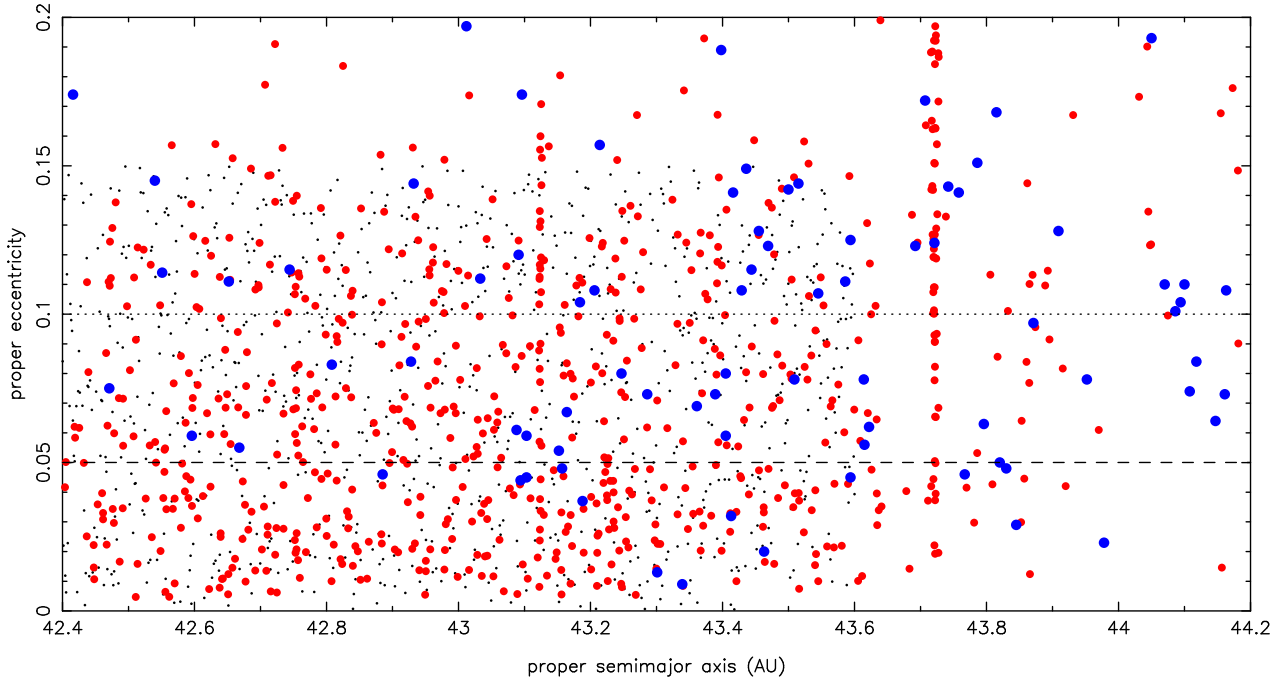}}
\caption{\small The same as Fig.~\ref{results} but for test particles initially with $a<43.6$~AU and $i>2^\circ$ and observed KBOs with current inclination larger than 4 degrees. Notice that the depletion of test particles with $e>0.05$ is much less pronounced than for the low-inclination population. 
}
\label{Hot}
\end{figure}

It can be immediately appreciated that the depletion of particles with
eccentricity above 0.05 is much less effective than for the cold
population. Thus, the hot population should have preserved its original
(i.e. post-capture) eccentricity distribution. For this reason, and
given the simple initial distribution (uniform) of our particles, the
simulated final distribution is not intended to match the
observations, but just to demonstrate that a moderate-eccentricity hot
population could have survived the same migration scenario that
explains the removal of the moderate-eccentricity cold population.

The differences between Fig.~\ref{results} and \ref{Hot} is so striking that it calls for an explanation. To understand what happens, we looked at the individual evolution of particles in both simulations. A representative evolution of particles removed at low inclination is shown in Figs.~\ref{lowi-evolve} and that of particles surviving at high inclination is shown in Fig.~\ref{highi-evolve}. 
Each figure shows the behavior of semi major axis (bottom panel) and eccentricity (top panel) over time, while depicting also the evolving location of the 4/7 and 5/9 resonances with Neptune. Notice that both particles have initially comparable values of semi major axis and eccentricity. The low inclination particle has $i=0.5^\circ$ and the high inclination particle has $i=19^\circ$.

\begin{figure}[t!]
\centerline{\includegraphics[height=12.cm]{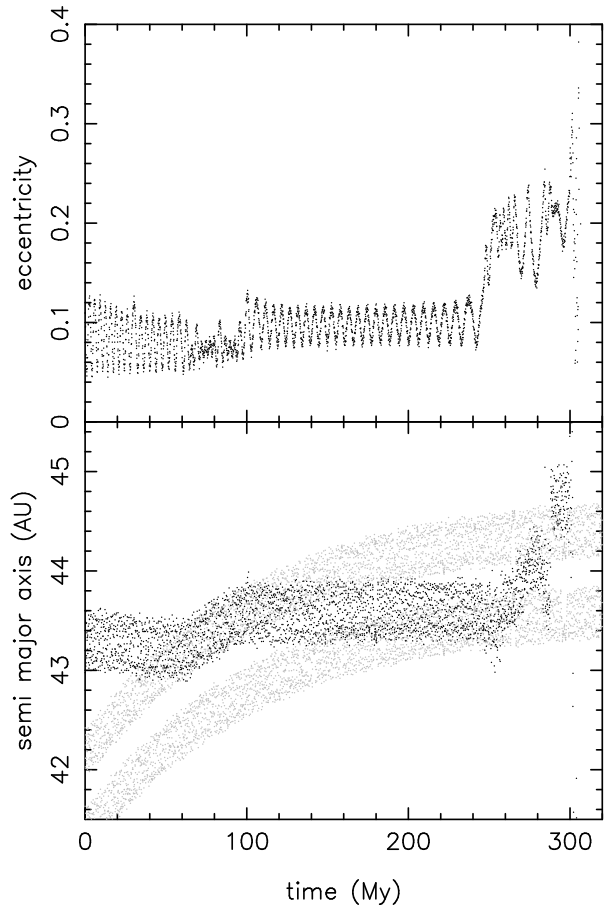}}
\caption{\small The evolution of a low inclination particle. The top
  panel depicts the time evolution of the eccentricity and the bottom
  panel that of the semi major axis (black dots). The gray  dots show
  the semi major axis of Neptune rescaled by the factors $(9/5)^{2/3}$ and 
$(7/4)^(2/3)$, namely they depict the locations of the 5/9 and 4/7 resonances.}
\label{lowi-evolve}
\end{figure}

Fig.~\ref{lowi-evolve} shows that the low-inclination particle was captured in mean motion resonance with Neptune twice. First, the particle was captured in the 5/9 resonance, between 60 and 100 My. This is clear from the drift in particle's semi major axis along the resonant track. Then the particle was released and it was captured in the 4/7 resonance at $t\sim 245$~My. This was a brief capture episode, but enough to raise the the eccentricity up to 0.2. Thus, the particle was destabilized: it was scattered by Neptune until it was dynamically removed. This shows that the cold population with moderate eccentricities was not lifted to into the hot population, but rather it was removed into the scattered disk. 

\begin{figure}[t!]
\centerline{\includegraphics[height=12.cm]{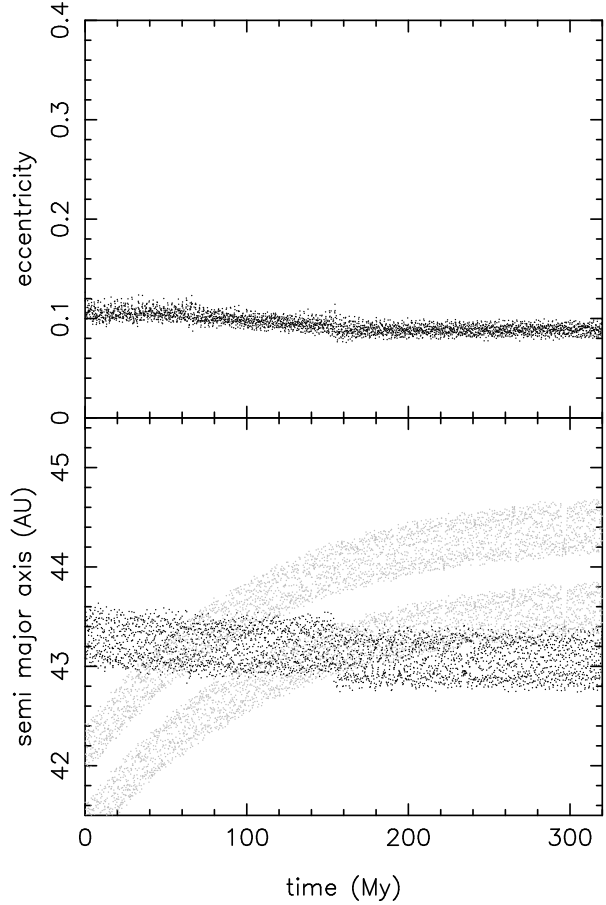}}
\caption{\small The same as Fig.~\ref{lowi-evolve} but for a high inclination particle.}
\label{highi-evolve}
\end{figure}

Fig.~\ref{highi-evolve}, shows a completely different behavior: the high-inclination particle crossed both resonances without being captured. The semi major axis and the eccentricity show a distinctive jump each time that a resonance was crossed, but the eccentricity was not significantly affected overall. Thus, the particle remained stable till the end of the simulation. 

Why is the probability of capture into resonance so different for low inclination and high inclination particles? We argue that the explanation is in the resonant structure. At small inclination, a mean motion resonance can be approximated by an integrable single-harmonic Hamiltonian. For the 
4/7 resonance the harmonic term is $e^3\cos(4\lambda_N-7\lambda+3\varpi)$ where $\lambda$ and $\lambda_N$ are the mean longitudes of the particle and of Neptune and $\varpi$ is the perihelion longitude of the particle. Similarly, for the 9/5 resonance the harmonic term is  $e^4\cos(5\lambda_N-9\lambda+4\varpi)$. But at high inclination there are additional major resonant harmonics. For the 4/7 resonance the second major harmonic is
$e i^2 \cos(4\lambda_N-7\lambda+\varpi+2\Omega)$, where $\Omega$ is the particle's longitude of node. For the 5/9 resonance one has 
two additional harmonics: $e^2 i^2 \cos(5\lambda_N-9\lambda+2\varpi+2\Omega)$ and $i^4\cos(5\lambda_N-9\lambda+4\Omega)$. 

As explained in Chapter 9 of Morbidelli (2002), a resonance described by multiple harmonics differing for the combinations of secular angles is analog to a {\it modulated pendulum}, which exhibits a wide chaotic band around the central resonant island. The probability of capture into resonance is very different in the integrable approximation and in the modulated pendulum approximation (Henrard and Henrard, 1991; Morbidelli and Henrard, 1993). In the case of the modulated pendulum, if the drift of the resonance is fast relative to the diffusion timescale inside the chaotic band, the capture into resonance is still possible (indeed we see in Fig.~\ref{Hot} particles aligned at the 4/7 resonance at $a\sim 43.7$ AU), but much more unlikely than in the integrable approximation. 

From this analysis we can also understand why the results for the cold population depend on the parameters $\tau$ and $e_0$ as reported in Table~1. If temporary resonant capture is the key, it is obvious that a faster migration speed (i.e. smaller $\tau$) makes less likely that particles are trapped in resonance. In the case of fast migration, most low-i particles just jump across resonance, with an evolution similar to that shown in Fig.~\ref{highi-evolve}. The dependence on eccentricity is more subtle. A larger Neptune's eccentricity makes the resonances more effective in exciting eccentricities through secular effects. In fact, if the eccentricity of the planet were null and the planet were not migrating, mean motion resonances would only force an eccentricity oscillation coupled with the resonant libration (see Morbidelli, 2002 - Chapter 9).  In the case of a migrating circular planet, particles -once trapped in resonance- would have their eccentricity increased monotonically as they move outward (e.g. see Malhotra, 1995 for an illustration concerning the 2/3 resonance). But in our case, this outward motion is short-ranged (less than $\sim 1$~AU), so this effect would not be very dramatic. The large and fast eccentricity increase observed in Fig.~\ref{lowi-evolve} at $\sim 250$~My is possible only because the eccentricity of Neptune is not null.    

It should be noticed in Fig.~\ref{Neptune} that the eccentricity of
the planet is not damped to zero in our simulations, so that the final
orbit of the planet is similar to the current one. This implies that,
in simulations starting with different initial eccentricities $e_0$,
eventually the eccentricity of the planet is the same. More precisely,
in the simulations with $e_0=0.05$ and $e_0=0.1$ the eccentricities of
Neptune became basically indistinguishable after 150My. Thus, why are
the resulting cold belts different, as suggested by the the data
reported in Table~1?.

\begin{figure}[t!]
\centerline{\includegraphics[height=9.cm]{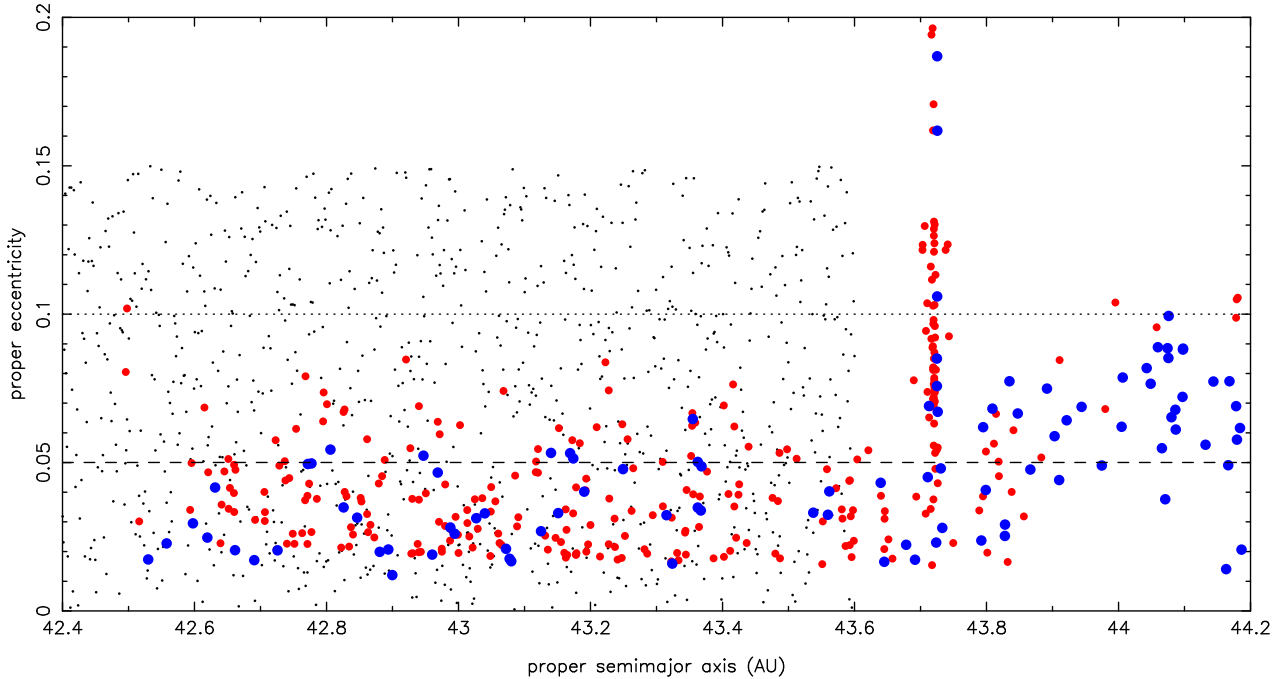}}
\caption{\small The same as Fig.~\ref{results} but for the simulation with initial Neptune's eccentricity $e_0=0.05$ (and still $\tau=$100~My). Notice that in this case the particle population initially extends only up to 43.6~AU.
}
\label{ae005}
\end{figure}

In the time range up to 150~My, the 4/7 resonance has swept the belt
up to 43~AU. The inspection of the final distribution of the particles
in the simulation with $e_0=0.05$ (Fig.~\ref{ae005}) shows that the
difference with the $e_0=0.1$ case is indeed mostly for particles with
$a\lesssim 43$~AU. Specifically, the $e_0=0.05$ simulations leaves
several particles there with $e>0.05$. whereas the $e_0=0.1$
simulation removed almost all of them. This difference is enough to
reduce drastically the probability to fit the observed distribution,
as reported in Table~1.

This explanation of the importance of the eccentricity of Neptune for the sculpting of the cold population has an interesting implication. It argues that the (already excellent) fit to observations obtained in the simulation with $\tau=100$~My and $e_0=0.01$ could improve if the eccentricity of Neptune were damped more slowly than we originally assumed. In fact, the number of particles surviving with $e>0.05$ and $a$ in the 43.0--43.5~AU range would presumably be reduced if Neptune's eccentricity remained somewhat larger until the 4/7 resonance reached 43.5~AU. 

 }

\section{Conclusions and discussion}
\label{conclusions}

The origin of the cold population of the Kuiper belt is still elusive.
A debate is open on whether this population formed in-situ or was
transported into the Kuiper belt from a location closer to the Sun
during the primordial ``wild'' evolution of the giant planet orbits.

Dawson and Murray-Clay (2012) pointed out a property of the cold
population that had passed previously unnoticed: the inner part of
this population (the one with $a<43.5$~AU) has eccentricities smaller
than $\sim 0.05$, despite orbits up to $e\sim 0.1$ could be stable in
this region. They interpreted this as an evidence that the cold
population was only moderately excited from its original circular
orbits, which argues in favor of its in-situ formation.

In this work we have confirmed the analysis of Dawson and Murray Clay
using proper elements. However, we showed that a slow migration of
Neptune (on a timescale of 100~My), initially on a moderately
eccentric orbit ($e\sim 0.075$), can reproduce the observations
starting from a population uniformly distributed in
eccentricity. 

Therefore, we disagree that the eccentricity distribution of the inner
cold belt can be used as an argument for minor excitation and in-situ
formation. Obviously, however, our results do not imply that the cold
population formed elsewhere and was transported into the Kuiper belt.
It's origin, therefore, remains elusive. 

On this subject, notice from Fig.~\ref{Helton} that the proper
eccentricity distribution of the cold population beyond the 4/7
resonance extends up to $\sim 0.1$.  It would be strange if the outer
population had been excited more than the inner population. This
suggests that the {\it entire} cold population got strongly excited,
so to cover the entire stability region (and probably going beyond
it), but it was then depleted near the stability border by the last
bit of resonance migration. We also notice in Fig.~\ref{Helton} that
beyond the 4/7 resonance there is a deficit of low-eccentricity
(i.e. proper $e<0.05$) cold objects, {known as the ``wedge''
  (Batygin et al., 2011)}. If this deficit is not due to observational
biases, {and it can not be explained the sweeping of the 5/9 resonance. Thus, it is an important diagnostic} feature for excitation/implantation models
of the cold population.

Our mechanism for the depletion of the inner cold population with
moderate eccentricities by resonance sweeping requires a migration
timescale of $\sim 100$~My.  This timescale is reasonable for a
tail-end of the planet migration process, and it is typically observed
in simulations where Neptune, after a wild phase of evolution due to
close encounters with the other planets, settles down in the
planetesimal disk (Gomes et al., 2005; Nesvorny and Morbidelli,
2012). The fact that our mechanism requires that the orbit of Neptune
had some moderate eccentricity when its semi major axis was at 28.5~AU
argues that the planet underwent previously some form of dynamical
instability. The reason is that planetesimal-driven migration can only
damp, {not excite,} the planet's eccentricity and therefore,
without an instability phase, the eccentricity of Neptune would have
always been small.

In principle, the eccentricity of Neptune at 28.5~AU could be the
remnant of a much larger eccentricity acquired when the planet was
closer to the Sun and had close encounters with the other
planets. Therefore, the results of this paper are not inconsistent
with the scenario of Levison et al. (2008) on the origin of the Kuiper
belt, provided that the tail-end of Neptune's migration is slow enough
(which was not the case in Levison et al., who adopted $\tau=1$~My
throughout their simulation). However, we acknowledge that said
scenario seems to be inconsistent with the existence of wide binaries
in the cold population (Parker et al., 2012).

More recently, Nesvorny and Morbidelli (2012) explored the possibility
that the outer solar system contained initially an extra Neptune-mass
planet, which eventually was ejected during the giant planet
instability. In the simulations that reproduced the best the current
orbits of the planets, Neptune had an evolution much less wild than
that considered in Levison et al. (2008). Neptune migrated out quite
smoothly and had only a moderate eccentricity excitation when it
encountered the lost planet.  {It will be important to investigate whether  this kind of evolution
can transport the cold population into the Kuiper belt via the
mechanism of Levison and Morbidelli (2003).  This
transport mechanism would in principle preserve wide binaries, satisfying the
constraint discussed in Parker et al. (2012) and would also be
consistent with the results of this paper.}

\acknowledgments 

A.M. thanks German Holmholtz Alliance for
funding this research through their “Plan
etary Evolution and Life” programme.
D.N. thanks NASA's OPR program for supporting his work.  

%\begin{thebibliography}{}

\section{References}

\begin{itemize}

\item[--] Batygin, K., Brown, 
M.~E., Fraser, W.~C.\ 2011.\ Retention of a Primordial Cold Classical 
Kuiper Belt in an Instability-Driven Model of Solar System Formation.\ The 
Astrophysical Journal 738, 13. 
\item[--] Brown, M.~E.\ 2001.\ The 
Inclination Distribution of the Kuiper Belt.\ The Astronomical Journal 121, 
2804-2814. 
\item[--] Dawson, R.~I., 
Murray-Clay, R.\ 2012.\ Neptune's Wild Days: Constraints from the 
Eccentricity Distribution of the Classical Kuiper Belt.\ The Astrophysical 
Journal 750, 43. 
\item[--] Duncan, M.~J., Levison, 
H.~F., Budd, S.~M.\ 1995.\ The Dynamical Structure of the Kuiper Belt.\ The 
Astronomical Journal 110, 3073. 
\item[--] Gladman, B., Marsden, 
B.~G., Vanlaerhoven, C.\ 2008.\ Nomenclature in the Outer Solar System.\ 
The Solar System Beyond Neptune 43-57. 
\item[--] Gomes, R.~S.\ 2003.\ The origin 
of the Kuiper Belt high-inclination population.\ Icarus 161, 404-418.
\item[--] Gomes, R., Levison, 
H.~F., Tsiganis, K., Morbidelli, A.\ 2005.\ Origin of the cataclysmic Late 
Heavy Bombardment period of the terrestrial planets.\ Nature 435, 466-469. 
\item[--] Henrard, J. and Henrard, M. 1991. Slow crossing of a stochastic layer. Physica D 54, 135-146.
\item[--] Kne{\v 
z}evi{\'c}, Z., Milani, A.\ 2000.\ Synthetic Proper Elements for Outer Main 
Belt Asteroids.\ Celestial Mechanics and Dynamical Astronomy 78, 17-46.  
\item[--] Kominami, J., Tanaka, 
H., Ida, S.\ 2005.\ Orbital evolution and accretion of protoplanets tidally 
interacting with a gas disk. I. Effects of interaction with planetesimals 
and other protoplanets.\ Icarus 178, 540-552. 
\item[--] Levison, H.~F., 
Morbidelli, A.\ 2003.\ The formation of the Kuiper belt by the outward 
transport of bodies during Neptune's migration.\ Nature 426, 419-421. 
\item[--] Levison, H.~F., 
Morbidelli, A., Van Laerhoven, C., Gomes, R., Tsiganis, K.\ 2008.\ Origin 
of the structure of the Kuiper belt during a dynamical instability in the 
orbits of Uranus and Neptune.\ Icarus 196, 258-273.
\item[--] Lykawka, P.~S., Mukai, T.\ 2005a.\ Long term dynamical evolution and classification of classical TNO$_{s}$.\ Earth Moon and Planets 97, 107-126.  
\item[--] Lykawka, P.~S., Mukai, T.\ 2005b.\ Exploring the 7:4 mean motion resonance I: Dynamical evolution of classical transneptunian objects.\ Planetary and Space Science 53, 1175-1187. 
\item[--] Malhotra, R.\ 1995.\ The 
Origin of Pluto's Orbit: Implications for the Solar System Beyond Neptune.\ 
The Astronomical Journal 110, 420. 
\item[--] Morbidelli A. and Henrard, J., 1993. Slow crossing of a stochastic layer. Physica D. 68, 187-200.
\item[--] Morbidelli A., 2002. Modern Celestial Mechanics. Taylor and Francis
\item[--] Morbidelli, A., 
Brown, M.~E.\ 2004.\ The kuiper belt and the primordial evolution of the 
solar system.\ Comets II 175-191. 
\item[--] Nesvorn{\'y}, D., Morbidelli, A.\ 2012.\ Statistical Study of the Early 
Solar System's Instability with Four, Five, and Six Giant Planets.\ The 
Astronomical Journal 144, 117. 
\item[--] Parker, A.~H., 
Kavelaars, J.~J., Petit, J.-M., Jones, L., Gladman, B., Parker, J.\ 2011.\ 
Characterization of Seven Ultra-wide Trans-Neptunian Binaries.\ The 
Astrophysical Journal 743, 1. 

\end{itemize}

%\end{thebibliography}
\end{document}